\documentclass{article}

\usepackage[nonatbib,preprint]{neurips_2019}
\usepackage[utf8]{inputenc} 
\usepackage[T1]{fontenc}    
\usepackage{hyperref}       
\usepackage{url}            
\usepackage{booktabs}       
\usepackage{amsfonts}       
\usepackage{nicefrac}       
\usepackage{microtype}      
\usepackage{graphicx}
\usepackage{multirow}

\newcommand{\eg}{\textit{e}.\textit{g}. }
\newcommand{\sectionsmallmargin}[1]{\vspace{-0.0cm}\section{#1}\vspace{-0.0cm}}
\newcommand{\paragraphsmallmargin}[1]{\vspace{-0.0cm}\paragraph{#1}}

\title{High Resolution Medical Image Analysis with Spatial Partitioning}

\author{
Le Hou$^{1,2}$ \quad Youlong Cheng$^{1}$\thanks{ylc@google.com} \quad Noam Shazeer$^{1}$ \quad Niki Parmar$^{1}$ \quad Yeqing Li$^{1}$\\
\textbf{
Panagiotis Korfiatis$^3$ \quad Travis M. Drucker$^4$ \quad Daniel J. Blezek$^3$  \quad Xiaodan Song$^{1}$ \
  \vspace{1mm}
}  \\
       $^1$Google Brain, Mountain View, CA \\
       $^2$Department of Computer Science, Stony Brook University, Stony Brook, NY \\
       $^3$Department of Radiology, Mayo Clinic, Rochester MN \\
       $^4$Enterprise Architecture, Mayo Clinic, Rochester MN
}

\begin{document}

\maketitle
\begin{abstract}
  Medical images such as 3D computerized tomography (CT) scans and pathology images, have $10^8$ to $10^9$ voxels/pixels. It is infeasible to train CNN models directly on such high resolution images, because of the memory limitation of a single GPU/TPU, and na\"ive data and model parallelism approaches do not work. Existing image analysis approaches alleviate this problem by cropping or down-sampling input images, which leads to complicated implementation and sub-optimal performance due to information loss. In this paper, we implement spatial partitioning, which internally distributes the input and output of convolutional layers across GPUs/TPUs. Our implementation is based on the Mesh-TensorFlow framework and the computation distribution is transparent to end users. With this technique, we train a 3D Unet on up to 512$\times$512$\times$512 resolution data. To the best of our knowledge, this is the first work for handling such high resolution images end-to-end.
\end{abstract}

\sectionsmallmargin{Introduction}
\label{sec:introduction}
Applying neural networks models \cite{ronneberger2015u,chen2017dcan,mobadersany2018predicting,lin2017focal} on high resolution image data, such as computerized tomography (CT) scans, satellite imagery, or histopathology slides is very computational extensive. For example, CT scans are typically acquired with sub-millimeter resolution resulting in image data sizes of 512$\times$512$\times$512 or more voxels. Furthermore, higher and higher resolution medical images are being collected \cite{ruddle2016design}. Assuming that neural activations are stored in half-precision floating point numbers (2 bytes) and the batch size is 8, a 1-layer Convolutional Neural Network (CNN) with 64 filters requires more than 137GB of GPU/TPU memory. To handle such high resolution input, existing models use a combination of down-sampling, dividing, and/or coarse-to-fine schemes \cite{hou2016patch,biswas2019smart,li2018h,vorontsov2018liver,chlebus2018deep}. An obvious drawback of these methods is that potentially useful information such as contextual features, small pathological volumes or high resolution details are lost.

To scale up neural network models, the data parallelism approach divides an input batch of instances into sub-batches and distribute them across multiple GPU/TPUs. Model parallelism \cite{dean2012large} approaches split and distribute model parameters, typically network layers, across multiple GPU/TPUs. These approaches do not solve the problem, since each GPU/TPU still need to process at least one high resolution image which results in more than 16GB of neural activations (in some cases $>$ 32GB) in a single GPU/TPU. More recent approaches \cite{jia2018beyond,shazeer2018mesh} are able to split the input batch along multiple dimensions, in addition to the batch dimension, and distribute parts across GPU/TPUs. However, it is not straightforward to split input images and distribute non-overlapping patches, since convolutions may take input across multiple patches. Splitting images into overlapping patches is not computationally efficient, because for most network architectures the overlap is so large such that almost the entire image is broadcast to every GPU/TPU. To overcome this, we implement spatial partitioning with halo exchange in tensorflow-TPU, to Mesh-TensorFlow. Halo exchange is a process during which GPU/TPU devices exchange data (patch margins) before convolution operations \cite{dryden2019improving}.

We evaluate our approach on the Liver Tumor Segmentation (LiTS) benchmark \cite{bilic2019liver} data (131 CT abdominal scans with liver segmentations). Directly training neural network models on high resolution images is prone to overfitting. To address this problem, we propose a synthesis-based data augmentation method for this application.

Our contributions are:
\begin{enumerate}
    \item An open source framework\footnote{Code: \url{github.com/tensorflow/mesh}. Example under  mesh/mesh\_tensorflow/experimental.} for training neural network models on high resolution images. It has the following advantages:
    \begin{enumerate}
        \item It supports training and evaluation on both GPUs and TPUs.
        \item It is highly efficient: spatial partitioning adds around 5\% to the total training time.
    \end{enumerate}
    To the best of our knowledge, we are the first to train neural networks on 512$\times$512$\times$512 resolution CT scans, in an end-to-end fashion.
    \item A data augmentation method for the liver tumor segmentation task using CT scans.
\end{enumerate}

\begin{figure}
  \centering
  \includegraphics[trim=20 150 20 100,clip,width=1.0\linewidth]{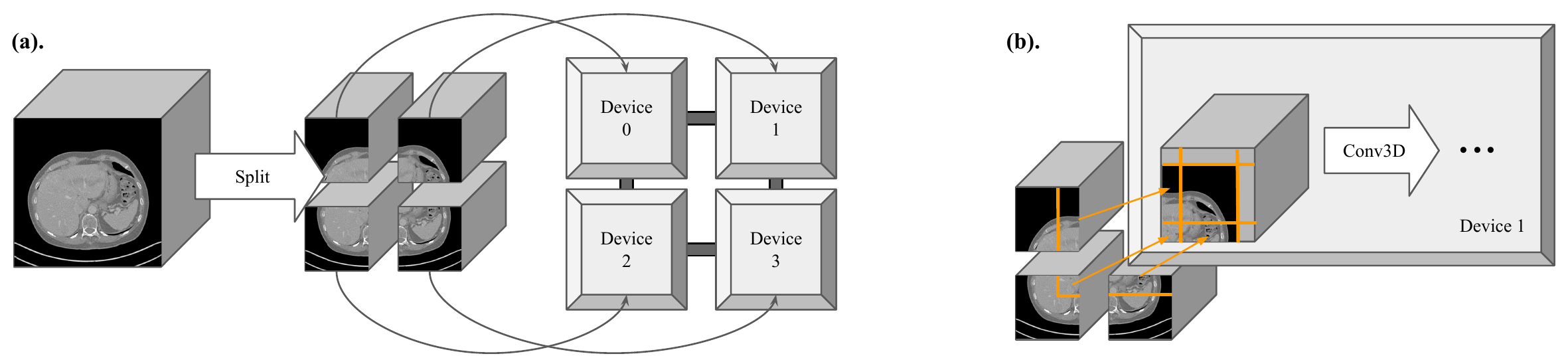}
  \caption{An illustration of spatial partitioning and halo exchange. \textbf{(a).} Spatial partitioning: we split very high resolution medical images into non-overlapping patches. Each computational device (GPU/TPU) processes one or more patches. \textbf{(b).} Halo exchange: before \textit{every} convolution operation, devices exchange patch margins (half the size of the convolution kernel) with each other.}
\label{fig:mtf-idea}
\end{figure}

\sectionsmallmargin{Mesh-TensorFlow with Spatial Partitioning and Halo Exchange}
Mesh-TensorFlow \cite{shazeer2018mesh} is a framework for large scale data and model parallelism. Given a collection of computational devices (\eg 8 GPUs), an end user defines how to map data dimensions to these devices. For example, splitting a batch of data along its batch dimension to 8 sub-batches and send each sub-batch to 1 GPU, is an 8-way data parallelism; splitting the parameters of a fully connected layer to 2 parts and send each part to 4 GPUs, is a 2-way model parallelism. Multiple such mappings can be defined for a model, and thus data and model parallelism can happen simultaneously. Mesh-TensorFlow internally transfers data across computational devices when necessary.

Mesh-TensorFlow is successful for training giant language models such as the transformer \cite{vaswani2017attention}. For image analysis tasks, high resolution images consume hundreds of gigabytes of GPU/TPU memory. To address this problem, one should split images along spatial dimensions to enable model parallelism in Mesh-TensorFlow. However, the current Mesh-TensorFlow framework does not support convolutional layers on spatially split images, due to the complexity and sliding-window nature of convolutions. Modeling a convolutional layer as many fully-connected layers in Mesh-TensorFlow is computationally unacceptable. 

To enable convolutional layers on spatially split images, splitting the input into overlapping patches is not computationally efficient, since the overlap might be very large, when the network is deep. We propose to simply exchange margins of patches across computational devices, then pad the patches with received margins, and finally apply convolution. We illustrate this process in Fig. \ref{fig:mtf-idea}. With this method, we are able to train a 3D Unet model \cite{ronneberger2015u,cciccek20163d} on 512$\times$512$\times$512 resolution CT scans.

\begin{figure}
  \centering
  \includegraphics[trim=5 145 5 152,clip,width=1.0\linewidth]{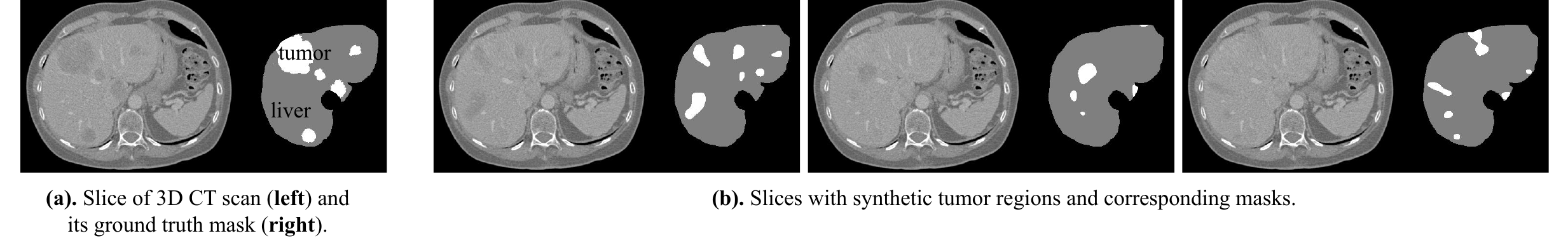}
  \caption{Given a real 3D CT scan with its ground truth mask (example of a 2d slice shown in \textbf{(a)}), we generate new training examples with corresponding masks (examples of 2d slices shown in \textbf{(b)}).}
\label{fig:synthetic}
\end{figure}

\paragraphsmallmargin{Data augmentation via Synthesizing}
\label{sec:synthesizing}
As opposed to training on 2D slices or 2D+ ``slabs'' of 3D scans, training on full resolution CT data directly is prone to overfitting, since there are fewer instances and more features per instance. To alleviate this problem, we propose a simple yet effective data augmentation method. We assume that on the training set, segmentation ground truth of liver and tumor are given. For each training image, we compute the intensity difference between tumor and non-tumor voxels in the liver region. We then ``remove'' tumor by subtracting the computed intensity difference. Finally, we synthesize tumor volumes by adding the computed intensity difference on random 3D volumes inside the liver. Boundaries around synthetic tumor are blurred (Fig. \ref{fig:synthetic}). Without this data augmentation method, the ``Dice per case'' scores (evaluation metrics are detailed in Sec. \ref{sec:experiments}) drop at least 10\%.

\begin{table}
  \caption{Validation results on the LiTS benchmark \cite{bilic2019liver}. We compute dice scores of 4 models and show their mean +/- standard deviation below. Note that the results are not directly comparable to the results on the LiTS challenge leader board, due to different evaluation metrics.}
  \label{tab:results}
  \centering
  \begin{tabular}{cccc}
    \toprule
    \multicolumn{2}{c}{Data resolution}  & Dice per case  & Dice global \\
    \midrule
    2D & 512$\times$512$\times$5 & 0.4072 +/- 0.0081  &  0.6432 +/- 0.0579 \\
    \midrule
    \multirow{4}{*}{3D} & 64$\times$64$\times$64  & 0.2513 +/- 0.0024  &  0.5364 +/- 0.0108 \\
     & 128$\times$128$\times$128 & 0.3589 +/- 0.0147  &  0.6494 +/- 0.0445 \\
     & 256$\times$256$\times$256 & 0.4359 +/- 0.0126  &  0.5783 +/- 0.1654 \\
     & 512$\times$512$\times$512 & \textbf{0.4547} +/- 0.0475  &  \textbf{0.7180} +/- 0.0446 \\
    \bottomrule
  \end{tabular}
\end{table}

\sectionsmallmargin{Experimental Results}
\label{sec:experiments}
We use the Liver Tumor Segmentation (LiTS) benchmark \cite{bilic2019liver} for the evaluation of our implementation. This dataset has 131 3D CT images. We randomly use 99 images for training and the remaining 32 for validation. We train 3D Unet \cite{cciccek20163d} models on four \textbf{3D resolutions: 64$^3$, 128$^3$, 256$^3$, and 512$^3$.} For the 64$^3$ resolution, we train a 3D Unet with 3 blocks (each block consists of 4 convolutional layers and 1 max-pooling) in the down-sampling (encoding) part. The up-sampling (decoding) part is symmetric to the down-sampling part. Each convolutional layer in the first block has 256 filters. The number of filters doubles after each max-pooling layer. For the $128^3$ resolution, we attach another block with 128 filters, right after the input layer. For the $256^3$ resolution, we further attach a block with 64 filters. Finally, for the 512$^3$ resolution, we attached a block with 32 filters. Thus networks at different resolutions have similar receptive field sizes. Since state-of-the-art methods on the LiTS benchmark work on 2D+ ``slabs'' instead of 3D data, we also test a 2D Unet on \textbf{512$\times$512$\times$5 resolution}. The 2D Unet has the same architecture as the 3D Unet except 3D convolution/max-pooling layers are changed to 2D. We apply data augmentation stated in Sec. \ref{sec:synthesizing}.

We train models on a cluster of TPUs. Each TPU has 2 cores. For $64^3$ and $128^3$ data, we use 2-way data parallelism and 16-way spatial partitioning on a TPU pod of 4$\times$4 TPUs. For 256$^3$ data, we use 2-way data parallelism and 128-way spatial partitioning on a TPU pod of 8$\times$16 TPUs. For 512$^3$ data, we use 2-way data parallelism and 256-way spatial partitioning on a TPU pod of 16$\times$16 TPUs. We use a batch size of 8 on 512$^3$ data, and 16 on other resolutions. We use the Adafactor \cite{shazeer2018adafactor} optimizer with a learning rate of 0.003. We use $0.9 \times Dice + 0.1 \times CrossEntropy$ as the loss function \cite{sudre2017generalised}.

\paragraphsmallmargin{Evaluation metrics}
We compute two dice scores as our evaluation criterion: dice per case, and dice global. For dice per case, we compute the dice score per validation image, then average the scores across all 32 validation scans. For dice global, we combine all scans as if there was only one scan and compute a dice score on the whole volume \cite{bilic2019liver}. We observe that the dice score fluctuates for the same network trained with different random initialization and different number of training iterations. Thus, we average the dice scores of 4 models (randomly initialize 2 times, then evaluate each model stopped at 2 different numbers of iterations).

\paragraphsmallmargin{Results and discussion}
From the results in Tab. \ref{tab:results}, we conclude that higher resolution data yields better Dice scores. Note that the results are not comparable to the results on the LiTS challenge leader board, due to different evaluation metrics. The Dice global scores have large standard deviations. This is because that a few CT scans in the validation set contain very large volumes of tumor, and hence prediction results of a few CT scans dominate the Dice global score.

\paragraphsmallmargin{Computational efficiency}
Our method is computationally efficient. In our experiments, operations introduced by spatial partitioning (partitioning, reshaping, and halo exchange) together add around 5\% to the total training time. In addition, more than 75\% of the total training time is spend on the forward and backward pass of convolutions.

\sectionsmallmargin{Conclusions}
It is challenging to train convolutional neural networks on high resolution medical images, since na\"ive data and model parallelism methods cannot effectively reduce the per-GPU/TPU memory requirements. We contributed a new Mesh-TensorFlow based framework which is capable of handling images of any size. To the best of our knowledge, we are the first to train neural networks on 512$\times$512$\times$512 resolution CT scans, without significant computational overhead.

{\small
\bibliographystyle{ieee}
\bibliography{main}
}

\end{document}